\documentstyle[epsfig,prl,twocolumn,aps]{revtex}

\catcode`@=11
\def\lsim{\mathchoice
  {\mathrel{\lower.8ex\hbox{$\displaystyle\buildrel<\over\sim$}}}
  {\mathrel{\lower.8ex\hbox{$\textstyle\buildrel<\over\sim$}}}
  {\mathrel{\lower.8ex\hbox{$\scriptstyle\buildrel<\over\sim$}}}
  {\mathrel{\lower.8ex\hbox{$\scriptscriptstyle\buildrel<\over\sim$}}} }
\def\gsim{\mathchoice
  {\mathrel{\lower.8ex\hbox{$\displaystyle\buildrel>\over\sim$}}}
  {\mathrel{\lower.8ex\hbox{$\textstyle\buildrel>\over\sim$}}}
  {\mathrel{\lower.8ex\hbox{$\scriptstyle\buildrel>\over\sim$}}}
 {\mathrel{\lower.8ex\hbox{$\scriptscriptstyle\buildrel>\over\sim$}}} }

\def\gsu{\raise4pt\hbox{\kern5pt\hbox{$\sim$}}\lower1pt\hbox{\kern-8pt
\hbox{$>$}}~}
\def\lsu{\raise4pt\hbox{\kern5pt\hbox{$\sim$}}\lower1pt\hbox{\kern-8pt
\hbox{$<$}}~}

\def\croce{\displaystyle / \kern-0.2truecm\hbox{$\backslash$}}
\def\lqua{\lower4pt\hbox{\kern5pt\hbox{$\sim$}}\raise1pt
\hbox{\kern-8pt\hbox{$<$}}~}
\def\gqua{\lower4pt\hbox{\kern5pt\hbox{$\sim$}}\raise1pt
\hbox{\kern-8pt\hbox{$>$}}~}
\def\mma{\lower1pt\hbox{\kern5pt\hbox{$\scriptstyle <$}}\raise2pt
\hbox{\kern-7pt\hbox{$\scriptstyle >$}}~}
\def\mmb{\lower1pt\hbox{\kern5pt\hbox{$\scriptstyle >$}}\raise2pt
\hbox{\kern-7pt\hbox{$\scriptstyle <$}}~}
\def\mmc{\lower4pt\hbox{\kern5pt\hbox{$<$}}\raise1pt
\hbox{\kern-8pt\hbox{$>$}}~}
\def\mmd{\lower4pt\hbox{\kern5pt\hbox{$>$}}\raise1pt
\hbox{\kern-8pt\hbox{$<$}}~}
\def\croce{\displaystyle / \kern-0.2truecm\hbox{$\backslash$}}
\def\quad@rato#1#2{{\vcenter{\vbox{
        \hrule height#2pt
        \hbox{\vrule width#2pt height#1pt \kern#1pt \vrule width#2pt}
        \hrule height#2pt} }}}
\def\quadratello{\mathchoice
\quad@rato5{.5}\quad@rato5{.5}\quad@rato{3.5}{.35}\quad@rato{2.5}{.25} }
\font\s@=cmss10\font\s@b=cmbx8
\def\reali{{\hbox{\s@ l\kern-.5mm R}}}
\def\m{{\hbox{\s@ l\kern-.5mm M}}}
\def\k{{\hbox{\s@ l\kern-.5mm K}}}
\def\naturali{{\hbox{\s@ l\kern-.5mm N}}}
\def\interi{{\mathchoice
 {\hbox{\s@ Z\kern-1.5mm Z}}
 {\hbox{\s@ Z\kern-1.5mm Z}}
 {\hbox{{\s@b Z\kern-1.2mm Z}}}
 {\hbox{{\s@b Z\kern-1.2mm Z}}}  }}
\def\complessi{{\hbox{\s@ C\kern-1.7mm\raise.6mm\hbox{\s@b l}\kern.8mm}}}
\def\toro{{\hbox{\s@ T\kern-1.9mm T}}}
\def\unity{{\hbox{\s@ 1\kern-.8mm l}}}
\font\bold@mit=cmmib10
\def\setbmit{\textfont1=\bold@mit}
\def\bmit#1{\hbox{\textfont1=\bold@mit$#1$}}
\catcode`@=12

\def\Ai{\hbox{\hbox{${\cal A}$}}\kern-1.9mm{\hbox{${/}$}}}
\def\Vi{\hbox{\hbox{${\cal V}$}}\kern-1.9mm{\hbox{${/}$}}}
\def\Di{\hbox{\hbox{${\cal D}$}}\kern-1.9mm{\hbox{${/}$}}}
\def\lam{\hbox{\hbox{${\lambda}$}}\kern-1.6mm{\hbox{${/}$}}}
\def\D{\hbox{\hbox{${D}$}}\kern-1.9mm{\hbox{${/}$}}}
\def\A{\hbox{\hbox{${A}$}}\kern-1.8mm{\hbox{${/}$}}}
\def\V{\hbox{\hbox{${V}$}}\kern-1.9mm{\hbox{${/}$}}}
\def\parz{\hbox{\hbox{${\partial}$}}\kern-1.7mm{\hbox{${/}$}}}
\def\B{\hbox{\hbox{${B}$}}\kern-1.7mm{\hbox{${/}$}}}
\def\R{\hbox{\hbox{${R}$}}\kern-1.7mm{\hbox{${/}$}}}
\def\si{\hbox{\hbox{${\xi}$}}\kern-1.7mm{\hbox{${/}$}}}

\begin{document}
\draft

\twocolumn[\hsize\textwidth\columnwidth\hsize\csname @twocolumnfalse\endcsname

\title{Metal-insulator crossover in superconducting  cuprates in  strong magnetic fields}
\author{P.A. Marchetti,$^a$   Zhao-Bin Su,$^b$ Lu Yu$^{c,b}$}

\address{ $^a$ Dipartimento di Fisica ``G. Galilei",INFN, I--35131 Padova, Italy\\
$^d$ Institute of Theoretical Physics, CAS, Beijing 100080, China\\
$^c$ Abdus Salam International Centre for Theoretical Physics, I-34100 Trieste, Italy}

\maketitle

\begin{abstract}
The metal-insulator crossover of the  in-plane resistivity upon 
temperature decrease, recently observed in several
classes of cuprate superconductors, when a strong magnetic
field suppresses the superconductivity, is  explained using 
the  $U(1)\times SU(2)$ Chern-Simons gauge  field theory.
The origin of this crossover is the same as that for a similar 
phenomenon observed in heavily underdoped cuprates without magnetic field.
It is due to the interplay between the diffusive motion of the charge carriers 
and the ``peculiar'' localization effect due to  short-range antiferromagnetic order.
We also calculate the in-plane transverse magnetoresistance
which is in a fairly good agreement with  available experimental data.
\end{abstract}

]

\narrowtext

\vskip 0.5truecm
The in-plane resistivity in heavily underdoped samples of cuprates
(in particular LSCO) exhibits a minimum   and a crossover from metallic to
insulating  behavior upon the temperature decrease\cite{takagi}.
Recently  a similar   crossover  was observed  in several classes of 
superconducting cuprates\cite{ando,four,ono,segawa} 
when  a strong magnetic field   (up to 60 Tesla)  suppresses 
the superconductivity.  The ``obvious'' interpretation in terms of
two-dimensional (2D)  localization or 2D insulator-superconductor
transition is ruled out, as  the sheet resistance, 
defined as $\rho_{sh} \equiv \rho_{ab}/a$, where $a$ is the interlayer distance,
at the crossover  point is between 1/25 to 1/12 in units of $h/e^2$\cite{ando,four,ono},
 or, using a free electron model,   the estimated product $k_Fl$,
where $k_F$ is the Fermi momentum, $l$ the mean free path,  is between
12 and 25! The insulating ground state persists up to
optimal doping in LSCO\cite{ando} and in electron-doped superconductors
Pr$_{2-x}$Ce$_x$CuO$_4$ (PCCO)\cite{four}, while in newly studied
Bi$_2$Sr$_{2-x}$La$_x$CuO$_{6+\delta}$ (BSLCO, or La-doped Bi-2201)
it persists only up to 1/8 hole-doping without showing any signature of
stripe formation. Thus ascribing this metal-insulator (MI) crossover to
a quantum critical point related to charge density instability\cite{caste}
is open to objections\cite{chakra}.
There were several attempts to  interpret the  insulating ground state
 in doped cuprates  using various
arguments on non-Fermi liquid behavior\cite{and}. However, the crossover
phenomenon  as temperature varies has not been addressed up to now, to the best of our
knowledge. It was thought earlier that the MI  crossover in the absence
of magnetic field\cite{takagi} and that induced  by  magnetic field is of 
different origin (the sheet resistance in the first case was substantially higher)\cite{ando}. 
This is doubtful, because the recent experiments on YBCO 
show a MI crossover in nonsuperconducting compounds in the absence of magnetic field
\cite{ando1}, and the same type of MI crossover (with comparable sheet resistance
at the crossover) in the superconducting compounds of the same series doped with Zn in the  
presence of magnetic field\cite{segawa}. We believe the {\it two crossover phenomena are of the
same origin}. We have used  the $U(1) \times 
SU(2)$ Chern--Simons (C.S.) approach to the $t-J$ model, proposed by us earlier
\cite{mar1}  to explain the MI crossover in heavily underdoped
cuprates in the absence of  magnetic field\cite{mar2}. 
In this Letter we  generalize
our formalism to include the effect of  magnetic field and show 
 that such a  MI crossover is a universal
feature of doped cuprates, and it is due to a ``peculiar charge localization'' effect
(using the wording of Ref. \cite{ono}), resulting from the interplay of the spin-excitation gap
(corresponding to short-range antiferromagnetic order (SRAFO)) and the holon
induced anomalous dissipation. Moreover, we will show that the observed 
large positive in-plane transverse magnetoresistance (MR) at low temperatures
\cite{kimura,lacerda} can be semi-quantitatively explained within this formalism.

The  $U(1) \times SU(2)$ C.S. gauge field approach is a particular scheme
of slave-particle formalism to treat the $t-J$ model based on an exact identity
\cite{fro1}, introducing a $U(1)$ field gauging the global charge symmetry
and a $SU(2)$ field gauging the global spin symmetry, both with C.S. actions. 
Using an optimization procedure of free energy\cite{mar1}, 
a careful mean field (MF) approximation gives the following results:
The $U(1)$ gauge field for low doping $\delta$ develops  a $\pi$-flux per plaquette
converting holons into Dirac fermions  with a Fermi energy $\epsilon_F 
\sim t \delta$. The holons induce a vortex structure in the MF 
configurations of the $SU(2)$ gauge field, with vortices centered at the 
holon positions. These dressed holons in turn are seen as slowly moving 
impurities  by  spin waves giving rise to  a spinon mass $m_s \sim 
\sqrt{\delta |\ln \delta|}$. Notice that this feedback is self-consistent
because for low  $\delta$ we have $\epsilon_F \sim t \delta << \epsilon_s
\sim J \sqrt{\delta |\ln \delta|}$. We use  $J = 0.1 eV, \; t/J = 3 $ in
our numerical computations. Due to a special choice of ``gauge
fixing'' (using the N\'eel gauge) the $SU(2)$ gauge field becomes physical,
describing the spin fluctuations.  The spinon action is given  by a nonlinear
$\sigma$-model with a mass term (spinon gap) which in the CP$^1$ representation
yields a new self-generated $U(1)$ gauge field $A$ coupling holons and spinons
( This field is analogous to the $U(1)$ gauge field 
in the standard slave-particle approaches\cite{ln,iw}).
Due to coupling to holons (fermions in our approach), this gauge
field acquires an anomalous dissipation term, the ``Reizer singularity''
\cite{reiz}, which dominates the low-energy action for the transverse
component of the gauge field $A^T$. For
$\omega, |\vec q|, \omega/|\vec q| \sim 0$ we have
$\langle A^T A^T \rangle (\omega, \vec q) \sim  
(\chi |\vec q|^2 + i\kappa {\omega \over |\vec q|})^{-1}$,
where $\chi \sim t/\delta$ is the diamagnetic susceptibility and 
$\kappa\sim \delta$ the Landau damping. The interplay of the two
different energy scales, the spinon gap and the holon
induced anomalous dissipation, is the key factor in our interpretation
of the MI crossover\cite{mar2}.
For the temporal component in the same limit, we have 
$\langle A^0 A^0 \rangle ( \omega, \vec{q} ) \sim ( \gamma + \omega_p )^{-1}$,
where $\gamma$ is the fermion density of states and $\omega_p$ the plasmon gap.

Now we consider the introduction of a magnetic field $H$ perpendicular to 
the plane. The Ioffe--Larkin rule \cite{il}, $R = R_s+R_h$, {\it i.e.}, the observed  resistivity is the sum
of the holon and spinon contributions,  can be generalized to this case.
The external electromagnetic  potential $A_{e.m.}$, corresponding to 
the constant magnetic field $H$ can couple with coefficient $-\varepsilon$ 
to spinons and $1-\varepsilon$  to holons. In principle, $ 0 \leq \varepsilon \leq 1$
is arbitrary.  However, to be consistent with the requirement that the 
physical inverse magnetitic susceptibility should be the sum of the inverse of 
that of holons and spinons, {\it i.e.}, $\chi^{-1} = (\chi^*_s)^{-1} + (\chi^*_h)^{-1}$,
where $\chi^*_s$ and $\chi^*_h$ are the renormalized spinon and holon susceptibility,
respectively  (this relation can be derived in the same way, as the Ioffe-Larkin rule),
we find $ \varepsilon = \chi^*_h/(\chi^*_h + \chi^*_s)$.
This value was argued earlier using  variational considerations\cite{ik}. 
Replacing these quantities by unrenormalized values, we find 
$1-\varepsilon \sim \chi_s/(\chi_h + \chi_s)
\sim {J\over t} \sqrt{\delta/|{\rm ln} \delta|} << 1$.
In the Coulomb gauge $A_{e.m.}^0 =0$,  the effective action for the gauge field 
$A$ can be written as:
\begin{eqnarray}
& &{\cal S}_{eff} (A)  = \int  dx^0 d^2x \left[ { i \over 2} [ A^0 (\Pi_{h}^0
 + \Pi_{s}^0) A^0 \right.\nonumber\\
 & & + (A^T+ (1-\varepsilon) A_{e.m.}) \Pi_h^{\perp} (A^T + (1-\varepsilon)A_{e.m.}) \nonumber\\
& &\left.+ (A^T - \varepsilon A_{e.m.})\Pi_s^{\perp} (A^T - \varepsilon A_{e.m.})]
+ { i \sigma_h(H) \over {2\pi}} A^0\epsilon_{ij} \partial^i A^j \right],\nonumber\\
 \end{eqnarray}
where $\Pi_h^{0,\perp}, \Pi_s^{0,\perp}$ are unrenormalized polarization 
bubbles  due to holons and spinons,
respectively, $ \sigma_h(H)$ is the  Hall conductivity due to holons. 
Note $A_{e.m.}$ appears in two places in this low energy 
effective action: one is simply a shift of the transverse component of the 
gauge field $A^T$ by $(1-\varepsilon)A_{e.m.}$ and $-\varepsilon A_{e.m.}$
corresponding to the minimal coupling to  holons and spinons, respectively,
while the other  is  a  C.S.  term due to parity breaking  induced by $H$.

As remarked  in \cite{ik}, the leading effect of the 
integration over $A_0$  is the renormalisation of the diamagnetic 
susceptibility: $\chi \rightarrow \chi (H) = \chi + {\sigma^2_h (H)\over 
4\pi^2 \gamma}$ in the $A^T$ effective  action.
The holon contribution $R_h$ can be evaluated using the  Boltzmann equation, taking into
account the classical cyclotron effect, as in\cite{ik}, obtaining
\begin{eqnarray}
& & R_h = R_h^0 [ 1+ (\frac{(1-\varepsilon)H\tau}{m_h})^2], \nonumber\\
& &R_h^0   \sim   \frac{m_h}{8} \frac{1}{\tau} \sim \delta 
\left[ \frac{1}{\epsilon_F \tau_{imp}} + ( \frac{T}{\epsilon_F} )^{4/3} 
\right],
\end{eqnarray}
where $\tau$ is the transport relaxation time, $\tau_{imp}$  the 
  impurity scattering time  and $m_h \sim  \delta/t $    the holon mass.
The spinon contribution $R_s$ is evaluated here using the Kubo formula for the spinon
current:
$ R_s = \lim_{\omega \to 0} \omega ( {\rm Im} \Pi^\perp_s (\omega) )^{-1} $, where
$\Pi^\perp_s$ denotes the transverse polarization bubble at $\vec{q} = 0$, 
renormalized by gauge fluctuations. At large scales, for $x^0 \gg |\vec{x}|$,
$\Pi^\perp_s(x)$ is approximately given by 
$\langle \partial_\mu G (x|A) \partial^\mu G (x|A) \rangle$, where
$\langle \; \rangle$ denotes the $A$-expectation value and $G(x|A)$ is 
the spinon propagator. Using the  Fradkin representation\cite{fradk,mar2}
it can be transformed into a gauge invariant form
$\langle \partial_\mu G (x|F) \partial^\mu G (x|F) \rangle$, where $G(x|F)$
in terms of a path-integral over 3-velocities, $\phi^\mu (t) \equiv \dot q_\mu (t),
\mu = 0,1,2$, is  given by
\begin{eqnarray}
& &G(x|F)  \sim  i \int^\infty_0 ds e^{-i m_s^2 s} \int {\cal D} \phi^\mu
e^{\frac{i}{4} \int^s_0 \phi^2_\mu(t) dt } \nonumber\\
& & \cdot \int d^3 p e^{i p x-ip^2s} e^{i {\cal Q}^{ij} (p,s,\phi|s^\prime, \lambda)
[F_{ij}( p, s|s^\prime, \lambda) + \epsilon_{i j} \varepsilon H]} 
\end{eqnarray}
with
\begin{eqnarray}
& & {\cal Q}^{ij} (p,s,\phi|s^\prime, \lambda)[\#] =
\int^1_0 d \lambda \lambda \int^s_0 d s^\prime \int^{s^\prime}_0 d 
s^{\prime \prime} \nonumber\\
& & \cdot ( \phi^i (s^\prime) - 2 p^i)(\phi^j (s^{\prime \prime}) -
2 p^j )[\#]
\end{eqnarray}
and
\begin{eqnarray}
& & F_{ij} ( p, s|s^\prime, \lambda) =
F_{ij}  (x + \lambda 
\int^{s^\prime}_0 d s^{\prime \prime} \phi(s^{\prime \prime}) - 2 p s),\nonumber\\
& & F_{ij}  \equiv  \partial_i A_j - \partial_j A_i.
\end{eqnarray}
After integration over $A^T$, using the effective action, and integration over 
the 3-velocities in the gaussian approximation, and over $p$ and  $s$ by
saddle point for low T, one obtains at large scales
\begin{eqnarray}
 \Pi_s^\perp (x) &\sim &\left[ \frac{\partial}{\partial x^\mu} 
\left[ e^{-i \sqrt{m^2-\frac{ T}{\chi (H)} f(\alpha) +
 \frac{\alpha^2 \varepsilon^2H^2}{3 q_0^2}}\sqrt{ x_0^2 - |\vec{x}|^2}} 
\right. \right.\nonumber\\
& & \cdot \left.\left. e^{-\frac{ T}{4 \chi(H)} q_0^2  g(\alpha) \frac{(x_0^2 - |\vec{x}|^2)}{m^2}}
\frac{1}{\sqrt{(x_0^2 - |\vec{x}|^2)}} \right] \right]^2,
\end{eqnarray}
where $\alpha = \frac{q_0 |\vec{x}|}{2}$, $q_0$ is the momentum scale
associated with the anomalous skin effect due to Reizer 
singularity: $q_0 \sim \bigl({\delta^2  T\over t}\bigr)^{1/3}$.  
$f$ and $g$ are functions describing  the effect of gauge fluctuations
and for a real argument, $f$ is  monotonically increasing, 
vanishing quadratically
near the origin and $g$ is monotonically decreasing vanishing at large
argument. (See \cite{mar3} for explicit expressions.) The integration over $|\vec{x}|$
and $x^0$ appearing  in the Fourier transformation are evaluated
by saddle point  for $|\vec{x}|$, at large $x^0$,  and
with scale renormalization by   principal  part evaluation for $x^0$\cite{mar2,mar3}. The integrals are 
dominated  by a complex saddle point at $|\vec{x}| =2 q_0^{-1} e^{i \frac{\pi}{4}}$
for $\chi(H) q_0|m_s(T, H)| \lsim T$, Im$( m^2_s (T, H)) \lsim m_s^2$,
where
\begin{equation}
m_s^2(T,H) = m_s^2 - i \left( \frac{c T}{\chi(H)} - \frac{\varepsilon^2H^2}{3 q_0^2} \right),
\end{equation}
with $c = -i f( e^{i \pi/ 4}$),  a constant with real part
$\sim 3$ and a small imaginary part. For the range of
physical parameters considered here ($H \lsim$ 100 Tesla), these bounds
gave a temperature range validity lying between a few tens and a few hundreds
degrees.

The saddle point produces the following effects: it induces  ``renormalization''
of the spinon mass yielding a $T$ and  $H$  dependent damping:
$m_s^2 \longrightarrow m_s^2(T,H)$; it gives rise to an attraction between
spinon and antispinon leading to the formation of a damped spin wave; 
it introduces a multiplicative renormalization of the correlation functions 
which for $R_s$ is given by 
$ Z(T,H) [ m_s^2(T,H)]^{\frac{1}{8}}$, where
$Z(T,H) = ( c^\prime  \frac{ T}{\chi(H)} q_0^{-3} 
- \frac{2}{3} \varepsilon^2 H^2 q_0^{-5})^{1/2}$  with $c^\prime$ a new constant
$\sim f^{\prime \prime} ( e^{i \pi/4})$. 

The final result in the range of $T$ described above is given by
\begin{equation}
R_s \sim Z(T, H) \frac{|m_s (T,H)|^{1/4}}{\sin \frac{\Theta (T,H)}{4}},
\end{equation}
where $m_s(T,H) \equiv |m_s (T,H)| e^{- i \Theta(T,H)}$.
Basic features of our formulas can be summarized as follows: for low $T$,
the effect of the spinon gap is dominating ($\Theta \searrow 0$), leading to
an  insulating behaviour; at higher temperatures one  finds a metallic
behavior due to the dissipation induced by gauge fluctuations, 
contained in $|m_s(T,H)|$, that becomes the dominant effect. Therefore
a MI crossover is recovered, decreasing the temperature.
The minimum  of $R$ as a function of $T$, $T_{MI} (\delta, H)$, is decreasing
with $\delta$ and increasing with $H$  (see the MR curve
$(R(H) - R(0))/R(0)$ in Fig. 1), 
in agreement with experiment\cite{segawa}. 
In the absence of magnetic field the crossover is determined by the interplay between 
$m_s^2 = \xi^{-2}$ and $T/\chi \sim T m_h \sim \lambda^{-2}$,
with $\xi$ the magnetic correlation length, $\lambda$ the thermal de Broglie wave
length. When $\lambda \le \xi$, the ``peculiar'' localization effect due to 
SRAFO is not ``felt'', and a metallic behavior is observed. In the opposite limit
 $\lambda \gg \xi$ the cuprate is insulating. The external magnetic
 field effectively reduces  the thermal energy, 
or increases the thermal wavelength, so the crossover temperature goes up.
The resistivity is diverging at
$T = \frac{\varepsilon^2H^2\chi(H)}{3c q_0^2} $,
which approaches $T=0$ as $H$ vanishes.
This divergence  is lying outside  the 
region of validity of our formulas and   should be considered as an 
artifact.  

\begin{figure}
\epsfxsize=2.8 in \centerline{\epsffile{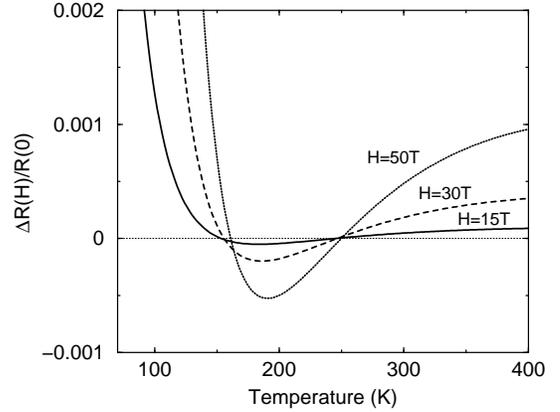}}
\caption[]{The calculated magnetoresistance for cases when the quantum effects
related to  $\sigma_h (H)$ are strong,  for  doping $\delta = 0.05$. 
It becomes negative near the minimum which itself
shifts to higher temperatures upon the field increase.}  
\label{fig1}
\end{figure}  

However,  the shift in MI crossover temperature   leads to a large positive (in--plane
transverse) MR at low $T$  which is our main new
result, and it was absent in the earlier treatments\cite{ik}.
The derived MR scales quadratically with $H$ (See Fig. 2) in agreement, 
in particular, with data on LSCO\cite{kimura,lacerda}, 
away from the doping $\delta = 1/8$ where the stripe effects dominate.  
As remarked in \cite{ik}, the shift of $\chi$ induced by the C.S.  term 
  reduces the damping and the $H^2$-term due to minimal coupling
acts in the same direction. In the region of $T$ where dissipation 
dominates this induces a reduction of resistivity, a tendency contrasted by 
the classical cyclotron effect. One then has two possible types of 
MR curves: one is always positive but it exhibits a knee 
below the crossover temperature between the mass gap and the dissipation 
dominated regions (See Fig. 3). This behavior can be compared with the one observed  in 
LSCO reported in \cite{lacerda}  and we find a reasonably good  agreement.
If, on the contrast,  the quantum effects related to  $\sigma_h (H)$ are  sufficiently strong,

\begin{figure}
\epsfxsize=2.8 in \centerline{\epsffile{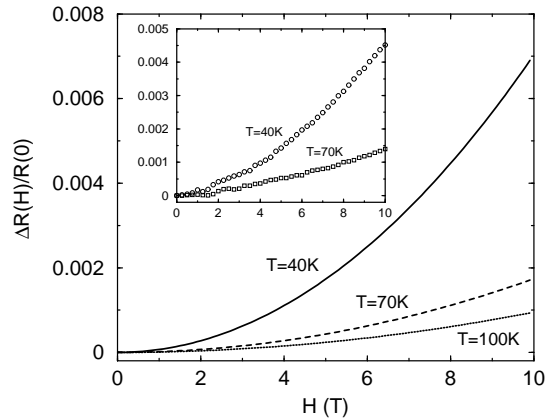}}
\caption[]{The calculated field dependence of the magnetoresistance
for doping $\delta = 0.075$, in comparison with experimental data on 
La$_{1.925}$Sr$_{0.075}$CuO$_{4+\epsilon}$ (inset), taken from Ref. \cite{lacerda}.}
\label{fig2}
\end{figure}

\begin{figure}
\epsfxsize=2.5 in \centerline{\epsffile{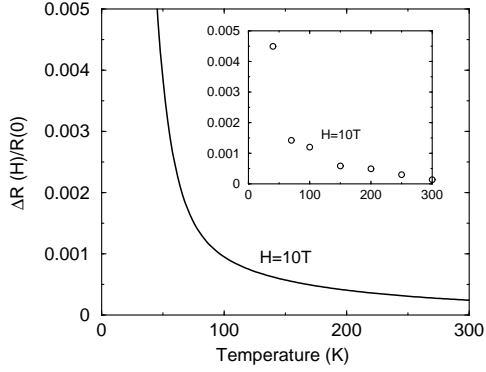}}
\caption[]{The calculated temperature  dependence of the magnetoresistance
for doping $\delta = 0.075$, in comparison with experimental data on 
La$_{1.925}$Sr$_{0.075}$CuO$_{4+\epsilon}$ (inset), taken from Ref. \cite{lacerda}.}
\label{fig3}
\end{figure}

\noindent  a minimum develops,  eventually leading to a negative MR   in  
some region around  it. This is illustrated in Fig.1. We should
point out that the large positive MR at low temperatures
is foreseen in this theory for  both cases.

A comment on the limit $H \sim 0$ is in order, where we recover the results of \cite{mar1},
in particular  $ m^2_s (T,0) = m^2_s - i cT/ \chi,\;  Z(T, 0) \sim \frac{1}{\sqrt{\delta}}$. 
In this limit the resistivity exhibits
an inflection point at temperature $T^* (\delta) \sim 200 - 300 $ K,
(found also in the experimental curves), above which the theoretical curves 
start to deviate strongly from the experimental data.
We propose to interpret this inflection point 
as a mid--point of a crossover to a new ``phase" where our MF treatment 
is not valid anymore. If we identify our $T^* (\delta)$ with the 
crossover temperature $T^*$ found in experiments, 
both  MI crossover temperature $T_{M I} (\delta) 
\equiv T_{MI} (\delta,  0) $ 
and $T^* (\delta)$  are found in a reasonable agreement with 
experimental data (in the
range $0.02 \lqua \delta \lqua 0.08$), 
due to a delicate cancellation of doping dependences: 
$\frac{T}{\chi m^2_s} \sim \frac{T \delta} { t \delta |\ln\delta|} = 
\frac{T}{ t \ln \delta}$.
$R_s$ in this limit can be written in terms of    a dimensionless
variable $x \equiv  c  T /\chi m^2_s$ apart from an  overall factor $\sqrt{|\ln{\delta}|}$.
Hence, if we neglect the contribution $\sim T^{4/3}$ due to holons and 
define a ``normalised resistivity" by $
\tilde R \equiv (R - R (T_{MI}))/(R (T^*) - R (T_{MI})),$
this is a function only of $ x $, thus exhibiting a 
``universal" behaviour, as observed in YBCO\cite{wuyts}.

As a  final remark it might be worthwhile to notice that the same $U(1) \times
SU(2)$ approach is able to reproduce qualitatively \cite{mar1,leo} the behavior of the 
spin lattice relaxation rate $(1/T_1 T)^{63}$ found in underdoped YBCO\cite{ber}
and a structure of Fermi arcs around $({\pi\over 2}, {\pi\over 2}$) in the 
spectral density detected by ARPES\cite{marsh}.

We thank J.H. Dai for collaboration in the early stage of this work, and 
Y. Ando  and T. Xiang for very helpful discussions.  The work of P. M.
is partially supported by RTN Programme HPRN-CT2000-00131, while L.Y. acknowledges
the partial support by INTAS Georgia 97-1340.

\end{document}